\documentclass[twocolumn]{aastex62}

\submitjournal{ApJ}

\shorttitle{Orphan GRB afterglow searches with the Pan-STARRS1 COSMOS survey}
\shortauthors{Huang et al.}

\begin{document}

\title{Orphan GRB afterglow searches with the Pan-STARRS1 COSMOS survey}

\author{Yun-Jing Huang}
\affil{Department of Physics, National Taiwan University - No.1, Sec.4 Roosevelt Road, Taipei 10617, Taiwan}

\author{Yuji Urata}
\affiliation{Institute of Astronomy, National Central University, Chung-Li 32054, Taiwan}

\author{Kuiyun Huang}
\affiliation{Center for General Education, Chung Yuan Christian University, Taoyuan 32023, Taiwan}

\author{Kuei-sheng Lee}
\affiliation{Department of Computer Science \& Information Engineering, National Central University, Chung-Li 32054, Taiwan}

\author{Meng-feng Tsai}
\affiliation{Department of Computer Science \& Information Engineering, National Central University, Chung-Li 32054, Taiwan}

\author{Yuji Shirasaki}
\affiliation{National Astronomical Observatory of Japan, 2-21-1 Osawa, Mitaka, Tokyo 181-8588, Japan}
\affiliation{Department of Astronomical Science, School of Physical Sciences, SOKENDAI (The Graduate University for Advanced Studies), 2-21-1 Osawa, Mitaka, Tokyo 181-8588, Japan}

\author{Marcin Sawicki}
\affiliation{Institute for Computational Astrophysics \& Department of Astronomy and Physics, Saint Mary's University, Halifax, Canada}

\author{Stephane Arnouts}
\affiliation{Aix Marseille Universit\'e, CNRS, Laboratoire d'Astro\-phy\-sique de Marseille, UMR 7326, F-13388,  Marseille, France}

\author{Thibaud Moutard}
\affiliation{Institute for Computational Astrophysics \& Department of Astronomy and Physics, Saint Mary's University, Halifax, Canada}

\author{Stephen Gwyn}
\affiliation{NRC-Herzberg, 5071 West Saanich Road, Victoria, British Columbia V9E 2E7, Canada}

\author{Wei-Hao Wang}
\affiliation{Academia Sinica Institute of Astronomy and Astrophysics, Taipei 106, Taiwan}

\author{Sebastien Foucaud}
\affiliation{Department of Astronomy, Shanghai Jiao Tong University, Dongchuan RD 800, 200240 Shanghai, China}

\author{Keiichi Asada}
\affiliation{Academia Sinica Institute of Astronomy and Astrophysics, Taipei 106, Taiwan}

\author{Mark E. Huber}
\affiliation{Institute for Astronomy, University of Hawaii at Manoa, Honolulu, HI 96822, USA}

\author{Richard Wainscoat}
\affiliation{Institute for Astronomy, University of Hawaii at Manoa, Honolulu, HI 96822, USA}

\author{Kenneth C. Chambers}
\affiliation{Institute for Astronomy, University of Hawaii at Manoa, Honolulu, HI 96822, USA}

%
\correspondingauthor{Yun-Jing Huang, Yuji Urata}
\email{yunjinghuang14@gmail.com, urata@g.ncu.edu.tw}

\begin{abstract}
We present the result of a search for orphan Gamma-Ray Burst (GRB) afterglows
in the Pan-STARRS1 (PS1) COSMOS survey. There is extensive theoretical and
observational evidence suggesting that GRBs are collimated jets; the direct 
observation of orphan GRB afterglows would further support this model. An optimal 
survey strategy is designed by coupling the PS1 survey with the 
Subaru/Hyper-Suprime-Cam (HSC) survey. The PS1 COSMOS survey, one of the survey 
fields in the PS1 Medium Deep Survey (PS1/MDS),
searches a field of 7 deg$^2$ from December 2011 to January 2014,
reaching a limiting magnitude R $\sim$ 23. The dense cadence of PS1/MDS is crucial 
for identifying transients, and the deep magnitude reached by the HSC survey (R 
$\sim$ 26) is important for evaluating potential GRB hosts. 
A transient classification method is employed to select potential orphan GRB 
afterglow candidates. After a thorough analysis of the transient and host galaxy 
properties, we conclude that there are no candidates in this survey field. 
The null result implies that the consideration of jet structures is essential for 
further orphan GRB afterglow surveys.

\end{abstract}

\keywords{Gamma-ray bursts (629), Transient sources (1851), Time domain astronomy (2109)}

\section{Introduction} \label{sec:intro}

Gamma-Ray Bursts (GRBs) are highly energetic explosions involving 
compact objects; they are caused by mergers or by the core 
collapse of massive stars \citep[e.g.][]{piran99}.
Jet collimation is needed to explain the large amount of isotropic 
equivalent energy released in the comparatively short prompt gamma-ray
phase of GRBs \citep[e.g.][]{sari99,frail01}. 
Furthermore, cocoon structures around ultra-relativistic jets are also
identified \citep{izzo19,chen20}.
Off-axis orphan afterglows (OAs) are a natural
consequence of GRB jet production (\citealt{rhoads1999}), and the
confirmed off-axis origin of X-ray flashes (XRFs) also
indicates the existence of OAs
\citep[e.g.][]{yamazaki02, granot02, granot05, urata}.
For both populations of the GRBs, short and long GRBs, unification along
with the GRB jet viewing angle is essential similar to the AGN model \citep{agn}.
As \citet{urata} verified XRFs as the off-axis viewing of long GRBs based
on both the prompt  emission (i.e. lower peak energy of prompt spectrum and 
energetic) and afterglow (i.e. multi-color brightening and supernova association) 
properties, further verification of off-axis viewing of GRB at the larger viewing 
angle (i.e. OAs) are crucial for unification of GRBs including related stellar 
explosions. In particular, off-axis viewing of classical short GRBs is essential to 
reveal the nature of short GRBs associated with gravitational wave transients caused 
by compact star merger \citep{2018MNRAS.476.4435L, 2017MNRAS.472.4953L, 2019MNRAS.489.1820L}. 
It is notable that most of theoretical models for 
GW170817/GRB170817A employed the complicated jet structure with off-axis viewing 
\citep[e.g.][]{alexander17, haggard17,lazzati17, murguia17,ioka18, jin18, 
kathirgamaraju18, troja18, troja19, 2018MNRAS.478..733L, 2018NatAs...2..751L, 2019ApJ...870L..15L}.

The mechanism of the collimated jet model is as follows: 
Relativistic matter with Lorentz factor $\Gamma$ is ejected as a
jet with opening half-angle $\theta_{jet}$. Radiation, on the other hand,
is beamed into a cone with opening angle $\Gamma^{-1}$, which is initially
inside the jet. 
Depending on the observation angle $\theta_{obs}$, the prompt emission shift to
the lower-frequency side or become invisible at $\theta_{obs} > \theta_{jet}$.
As $\Gamma$ decreases, the radiation cone spreads,
giving rise to an afterglow emission,
with wider angular range and fainter magnitude than the initial prompt emission. 
When $\Gamma^{-1}$ exceeds $\theta_{jet}$, 
it will cause two observable effects, dependent on the observation angle
$\theta_{obs}$: (1) achromatic breaks in the light curves for on-axis observers 
($\theta_{obs} < \theta_{jet}$); (2) the appearance of off-axis orphan afterglows
for off-axis observers ($\theta_{obs} > \theta_{jet}$).
Although chromatic temporal
afterglow evolution from X-ray to optical is one of the puzzles of
GRB physics, the observation of achromatic breaks in a number of GRB
afterglows supports the existence of jet collimation
(e.g. \citealt{harrison1999}); the finding of OAs
would provide additional direct observational evidence for it.

The event rate of OAs depends on the jet
structure.  \cite{N02} considered GRBs to have constant total energy
and a universal post-jet-break light curve, 
with jets having a constant maximal observing angle $\theta_{max}$ 
that is independent of $\theta_{jet}$ in the case of $\theta_{jet}
< \theta_{max}$, and derived the maximal
flux at $\theta_{obs}$ to estimate the event rate at the limiting
magnitude of observing instruments. 
\cite{T02} used average GRB parameters from a sample of 10 well-studied events,
and estimated the event rate in the framework of the collimated jet
model. Both of these studies
considered a uniform jet with sharp edges: the ``top-hat" model. 
\cite{R08}, on the other hand, considered 
a jet with a wide outflow angle
$\theta_{jet}$ = 90$^{\circ}$ and an angle-dependent energy
distribution E($\theta$) $\propto \theta^{-2}$, the universal structured
jet (USJ) model.
The predicted rates from
the three papers differ by about an order of magnitude for an
all-sky snapshot at a given observational sensitivity in optical range
(\citealt{R08}, Figure 8).  Therefore, systematic surveys for off-axis
OAs differentiate between the models by constraining the event rate.

Previous failed attempts at OAs searches have been 
numerous in various wavelengths: X-ray (e.g. \citealt{grindlay1999, greiner2000}), optical (e.g. \citealt{rau2006, malacrino2007})
and radio band (e.g. \citealt{levinson2002, galyam2006}).
\cite{grindlay1999} found 13 candidates in the Ariel 5 survey,
and set the all-sky event rate $\sim$0.15 day$^{-1}$,
which is consistent with BATSE.
\cite{greiner2000} found 23 candidates in the ROSAT all-sky survey,
but these were later shown to be mostly, if not entirely, from flare stars.
These results indicate that there is no marked difference between
the beaming angles of prompt gamma-ray and X-ray emissions. There have
been several optical surveys searching for OAs
using different sky coverage and observation depths. The
Deep Lens Survey transient search reached a sensitivity of 24 mag and
surveyed an area of 0.01 deg$^2$ yr
(\citealt{becker2004}). \cite{rykoff2005} surveyed a wide field of
effective coverage 1.74 deg$^2$ yr (but with a low sensitivity of 17.5
mag) using the ROTSE-III telescope. \cite{rau2006} used the 
Wide Field Imager (WFI) attached to the 2.2-m MPG/ESO telescope to
survey an area of 12
deg$^2$, with a sensitivity as low as R = 23 mag. They observed for 25
nights with one- or two-nights separation. \cite{malacrino2007}
performed a search using the CFHTLS very wide survey with a
sensitivity as low as R = 22.5 mag and an area of 490 deg$^2$. However,
all these attempts have failed to provide a firm detection of off-axis
OAs, null results are in agreement with theoretical
predictions \citep[e.g.][]{T02}.  Recently, \cite{Law18} reported
the discovery of a radio
transient that has properties similar to either a magnetic nebula or an
OA, with the evidences strongly suggesting the
latter; \cite{Marcote19}, by examining source properties, 
have later supported that this radio transient is likely an OA.

In this paper we report a systematic survey of OAs
using Pan-STARRS1.  The paper is structured as follows: In
$\S$\ref{[sec:stra]}, we describe our observational strategy: coupling the
Pan-STARRS1 and HSC surveys. In $\S$\ref{sec:obs}, we explain the details
of the instrumentation and the survey duration. In $\S$\ref{sec:analysis}, we
describe our transient classification method and our analysis of
photometric redshifts (zphot) using the Le Phare program \citep{lephare1, lephare2}, 
and present our result. In $\S$\ref{sec:discussion}, we discuss the predicted
detection rates from three theoretical papers and compare them with our
result; we also calculate OAs rates for
future prospective surveys such as HSC and LSST. Finally, we present
our conclusions in $\S$\ref{sec:conclusion}.

\section{Survey Strategy}\label{[sec:stra]}

\begin{figure*}
\begin{center}
	\includegraphics[width=1.8\columnwidth]{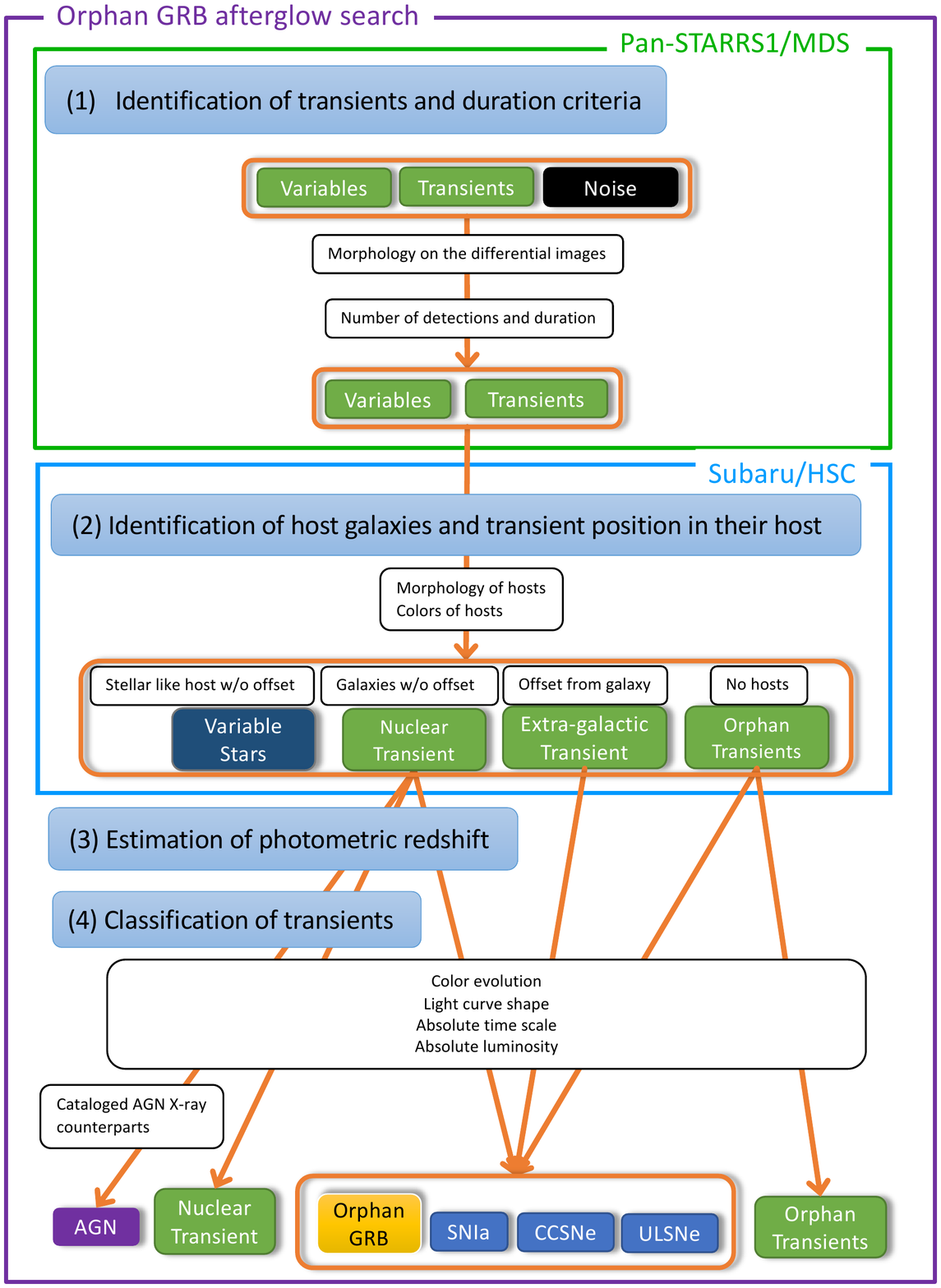}
\end{center}
\caption{Strategy of survey.}
\label{flowchart}
\end{figure*}

The expected properties of an OA are (1)
absence of prompt emissions in the high energy band; (2) brightness fainter
than that of on-axis GRB afterglows; (3) a light curve with three components
(rise, peak, and rapid decay); (4) the same
optical color as on-axis afterglows; and (5) association with a host
galaxy having properties similar to the host galaxies of on-axis GRBs. 
Taking these properties into account, we
designed OA searches using Pan-STARRS1
and Subaru/Hyper-Suprime-Cam (HSC). Our basic search-procedure
is shown in Figure~\ref{flowchart}. 
Similar to detections of prompt emission with the uniform spatial distributions of 
GRB on the celestial sphere, surveys using wide field of view (FOV) detectors with 
larger than several thousand square degrees such as {\it Swift}/BAT \citep[1.4 
steradian (half coded);][]{bat} and {\it HETE-2}/WXM 
\citep[$80^{\circ}\times80^{\circ}$;][]{wxm} are desired for efficient OA searches.
However, because of the long lifetimes of the afterglows,
the FOV of instruments can be reduced by using tiled observations 
with the appropriate cadence and pattern.
Therefore, untargeted transient surveys
with optical wide-field imagers, such as the PanSTARRS1 Medium Deep
Survey (MDS) and the Subaru/HSC, are sufficient for these searches.

One of the challenges for such generic optical transient surveys is
distinguishing OAs from other types of optical transient,
because candidates are expected to be rare compared
with supernovae (SN) of known types.
We designed a seven-step procedure for finding OAs: (1)
creating differential images using reference-stacked and nightly
stacked images, (2) generating light curves for transient components,
(3) identifying host galaxies, (4) measuring transient locations in
hosts, (5) matching with known sources in various catalogs, (6)
matching of light and color temporal evolution patterns, and (7)
estimating of the photometric redshifts of hosts.
These selections have also been providing other rare transient
phenomena \citep[e.g.][]{urata12,tanaka14,cenko15}.

We used Pan-STARRS1/MDS for finding transients and
characterizing their light curves and colors, and Subaru HSC for
identifying host galaxies and obtaining photometric redshifts.
The dense and continuous monitoring ability of the Pan-STARRS1/MDS fields 
is crucial, given the duration of the expected OAs.
Considering the limiting magnitudes of Pan-STARRS1/MDS,
the expected observable duration width is about one or two weeks,
relatively short compared to the typical variable time scale of AGN and SN.
The multi-color observing capability of Pan-STARRS1/MDS is crucial
for characterizing light curves and colors. The colors of classical
GRBs afterglows (e.g., $g$-$r$=0.5$\pm$0.2, $r$-$i$=0.6$\pm$0.2)
(\citealt{simon2001, li2018}) exhibit no temporal evolution
and no redshift dependency, because these emission mechanism of optical
afterglows, unlike that of other transients, is synchrotron radiation,
usually describable by a simple power-law. Given the limiting magnitude of
Pan-STARRS1/MDS, the expected redshift range of OAs is z $\sim$1.
The brightness range of host galaxies
for classical GRBs at $z$ up to $\sim$1 is from 23.0 to 25.5 mag in $r'$/R band
(e.g. \citealt{berger}), which Subaru/HSC images are deep enough to detect.

\section{Observation and Data} \label{sec:obs}

\subsection{Pan-STARRS1 Survey} 

Pan-STARRS1 (PS1) is a 1.8-meter telescope located at the summit of
Haleakala on the island of Maui, Hawaii. It performs a wide field
optical sky-survey with a field of view of 7 square degrees using a
mosaic CCD camera with sixty 4800$\times$4800-pixel detectors
(0.26$\arcsec$ per pixel). The full description of the system is given
in \cite{tonry2012} and \cite{ps1}. PS1 uses five broadband filters,
designated as $g_{P1},\ r_{P1},\ i_{P1},\ z_{P1}$, and $y_{P1}$. The
first four are similar to the SDSS filters
$g_{SDSS},\ r_{SDSS},\ i_{SDSS}$, and $z_{SDSS}$, but different in
that $g_{P1}$ extends 20 nm redward of $g_{SDSS}$ and $z_{P1}$ is cut
off at 920 nm. The range of $y_{P1}$ is roughly from 920 nm to 1050 nm.
Further information on the filter and photometric system is given in
\cite{tonry2012}.

The PS1 Medium Deep Survey (MDS) surveyed 10 fields, each
with an area of 7 square degrees. In this work we evaluate the MD04
field of PS1 MDS, which is centered at RA(J2000) =150.000$\degr$,
Dec(J2000) =2.200$\degr$, and overlaps
a well-studied field, the COSMOS field. 
We can make use of the extensive multi-band
data from other surveys to classify our transients. The cadence and
filter cycle are as follows: Each night 3-5 MD fields are observed,
using both $g_{P1}$ and $r_{P1}$ on the first night, $i_{P1}$ on the
following night, and $z_{P1}$ on the third. $y_{P1}$ is used around
full moon. The exposure times for each filter on each night are:
8$\times$113s for $g_{P1}$ and $r_{P1}$, and 8$\times$240s for the
rest. Each night the 8 exposures are dithered through the Image
Processing Pipeline (IPP) (\citealt{ipp}; \citealt{ipp2}) and combined
into nightly stacks of durations 904s and 1902s, producing a 5$\sigma$ 
depth of r $\sim$ 23.3. 

\begin{figure}
\epsscale{1.0}
\plotone{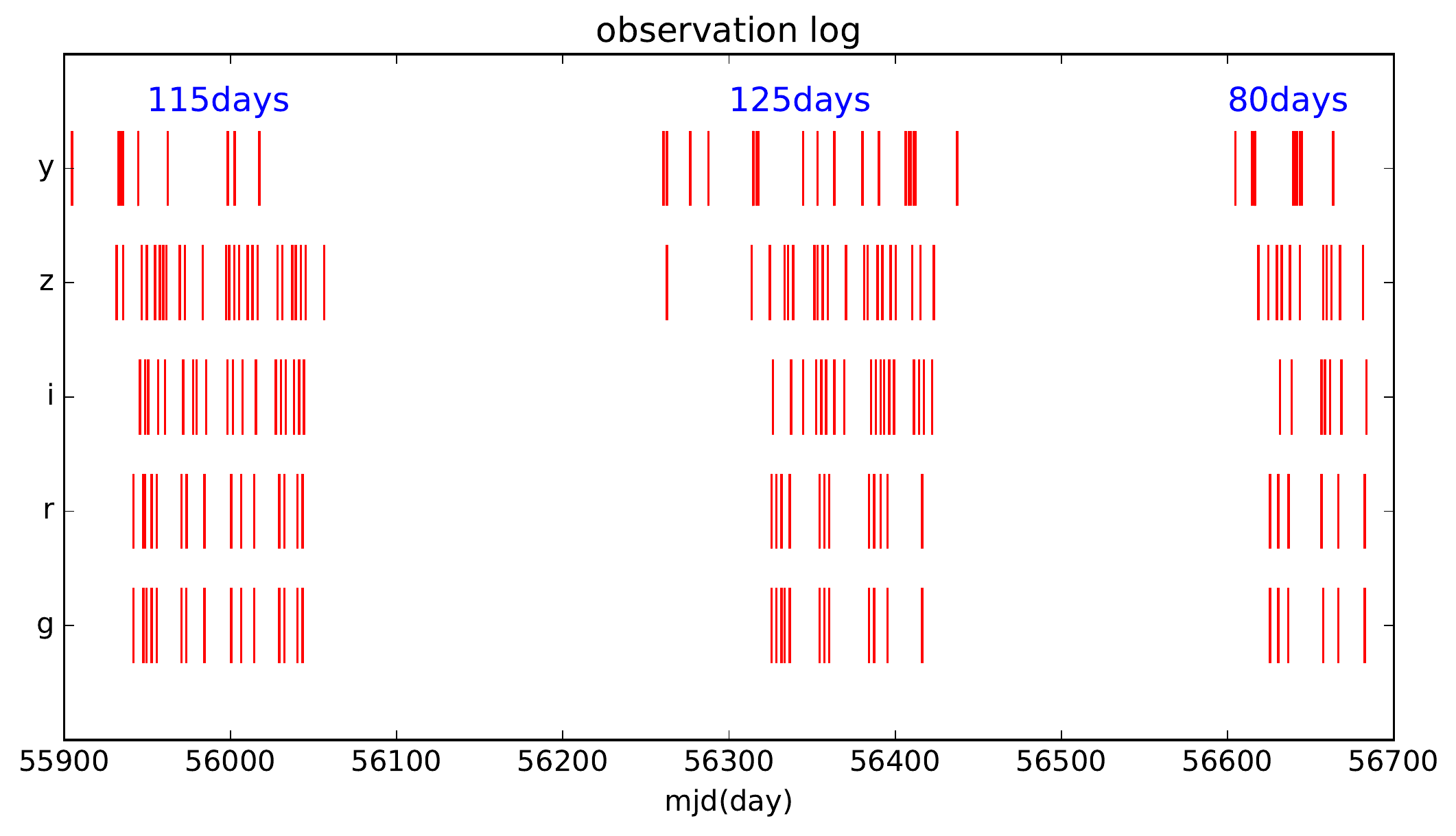}
\caption{Observation log of PS1 MD04 field. Each red vertical line marks the night 
observed. Each row shows the survey condition in different bands. The plot shows that
the survey is divided into three periods, the length of each is denoted by the blue 
text in the upper row.}\label{log}
\end{figure}

Survey cadence with coupling of filters are critical for selecting OA
candidates. The actual cadence of survey with individual filter for
MD04 are shown in Figure~\ref{log}, where
we can see that the survey is divided into three survey periods
over the span of $\sim$ 2 years (from December 2011 to January 2014).
Each survey period lasts about 3 to 4 months. 
There were 115 nights for the first year, 125 nights for the second year,
and 80 nights for the third year, resulting in a total of 320 nights.
To calculate the total effective OA survey term, we excluded the isolated
nights which is separated by more than 20 nights from other
observations, since these are only snapshots and cannot be used
to identify a transient. Thus, the exact number of effective OA survey
term is 154 nights. This effective survey term is used for calculation
of the expected OA rate in $\S$ \ref{sec:theory}.

The IPP for MDS image processing was originally located 
at the Maui High-Performance Computing Center (MHPCC), and is moved 
to the Information Technology Center at University of Hawaii.
It has several nightly processing stages: 
First, in the Chip Processing stage, the individual CCD
chips are detrended and sources are detected and characterized.
Then, in the Camera Calibration stage, the CCDs of each full exposure are calibrated. 
The images are later geometrically transformed into common pixel-grid images (called
skycells) in the Warp stage. These skycell images are then combined
to generate nightly images in the Stack stage. 
Next, source detection is performed in the Stack Photometry stage in all five
filter stack images at the same time. 
Convolved galaxy models are fitted in the Forced Galaxy Models stage. 
Finally, in the Difference Image stage, nightly stacks are 
compared to a template reference stack for MDS fields.

\subsection{SUBARU/HSC Observations}

Subaru is an 8.2 m telescope located at the summit of Mauna Kea,
Hawaii. The Hyper Suprime-Cam (HSC) (\citealt{hsc}, \citealt{hsc2}) is
a wide-field imaging camera with a field of view of diameter 1.5 deg
and 116 2048$\times$4096-pixel CCDs (0.168$\arcsec$ per pixel),
mounted on the prime focus of Subaru. Under the Hyper Suprime-Cam
Subaru Strategic Program (HSC-SSP) (\citealt{hscsurvey}),
three-layered (Wide, Deep, and Ultradeep), multi-band (griz plus
narrow-band filiters) imaging surveys have been executed, starting in
2014.
In this paper, we use the Deep and UltraDeep
data, which have survey fields overlapping the PS1 MD04 field:
E-COSMOS and COSMOS, respectively. E-COSMOS has four pointings in the
Deep layer, and overlaps COSMOS, which has one pointing in the UltraDeep layer.

Only six filters can be used in a single observing run, so typically
four or five broad-band filters plus one or two narrow-band filters
are used. The exposure times are as follows: For the Deep layer, a
single exposure lasts 3 min for the g and r bands, and 4.5 min for the i, z, and
y bands. For each field, 3-5 exposures are taken on each night in each
filter. For the UltraDeep layer, a single exposure lasts 5 min for all
broad bands, and 3-10 exposures are taken in one night. A more
complete description of the HSC survey is provided by
\cite{hscsurvey}.

Images are processed through the hscPipe pipeline, which consists of
four stages: The CCDs from each visit are calibrated in the CCD
Processing stage. Then observations from different visits are further
calibrated in the Joint Calibration stage. Subsequently, images
from different visits, including observations on different nights,
are combined into a deeper coadded image in
the Image Coaddition stage, which is further processed in the Coadd
Processing stage to detect and measure objects. The pipeline is fully
described in \cite{hscpipe}. 
The astrometry and photometry are calibrated against the PS1 PV2 catalog.
The process resulted in reduced data with an astrometirc accuracy of 30
mas and a photometric accuracy of $\sim2\%$ \citep{tanaka17}.
We use the s16a data release of HSC SSP,
which contains the data obtained from January to April 2016 processed
and merged with the s15b data release, that is, SSP data taken from
March 2014 to November 2015. The depths of images with 5$\sigma$
confidence level for point sources are r mag $\sim$ 27.1 (Deep) and
27.7 (Ultradeep).

\subsection{Complementary UV and NIR Data} \label{uir}
In order to optimize the photometric redshift estimation for the
hosts, complementary UV and NIR data are required. In addition to
Subaru/HSC data, we used near-infrared data (y, J, H, and Ks band)
from the UltraVista data release 3 (\citealt{ultravista}), and U-band data
from MUSUBI (W.-H. Wang et al., in prep.) and CLAUDS \citep{clauds} to generate
the spectral energy distribution (SED) of
hosts, and compute their photometric redshifts. The photometry used is
aperture 2 magnitude. MUSUBI, which stands for Megacam Ultradeep
Survey with U-Band Imaging, is a U-band complementary data set for the HSC
UltraDeep layer. CLAUDS, which stands for CFHT Large-Area U-band Deep Survey,
is a U-band complementary data set for the HSC Deep layer. 
MegaCam has two U-band filters u* and u \citep{clauds}.
In this survey, the data from both MUSUBI and CLAUDS used the u* filter.

\section{Result and Analysis} \label{sec:analysis}

\subsection{Overview of Analysis}

OA searches were performed based on the
flowchart shown in Figure~\ref{flowchart}. Here, we describe the four steps
of our analysis: (1) identification of transients with durations
shorter than 15 days from the PS1-MD04 data ($\S$ \ref{short
  duration}); (2) identification of host galaxies and assessment of
transient positions within them using the Subaru/HSC survey
($\S$ \ref{host}) and available galaxies catalogues; (3) estimation of
photometric redshift ($\S$ \ref{photometric redshift}); and (4)
classification of transients ($\S$ \ref{sec:results}).
Utilizing these results,
we examine whether the transients have the
expected properties of OAs.

\subsection{Identification of Transients}\label{short duration}

For transient identification, we used the difference catalog produced by
the PS1 Transient Science Server (TSS) (\citealt{tss, mccrum}). TSS creates 
difference images by comparing nightly stacks
made by IPP compared to manually created reference images. Point
spread function (PSF) fitting photometry is performed on the
difference images and catalogs of transients are produced.

We imported individually detected transient candidates from each difference catalog
into a custom-made PostgreSQL database and
performed cross-matching with their locations. For the cross-matching,
we set a search radius of 1$\arcsec$ for matching identical
objects, and assigned a {\bf count number} representing the number of
multiple detections with different filters and/or epochs, which is useful to exclude 
noise events.  The location of each multi-detection transient was
taken to be the average location of the individual detections. In total,
136657 transient candidates were identified in the PS1-MD04 field
after importing all difference catalogs made during the 2 years of the
survey.

We selected short duration candidates by applying two criteria:
(selection-A) count number $\geq$ 3 and observed duration $<$ 15 days;
(selection-B) count number = 2, observed duration $<$ 4 days, and a
decaying light curve. The duration cut for the count = 2 case 
was chosen to be shorter on the grounds that,
since the rising phase is more rapid than the decaying phase,
if an OA can only be detected twice, 
its rising phase is too rapid to be detected, and only its
rapid decaying phase is detectable.
Transients with count = 1 are not considered for candidate selection 
because they are more likely to be noise. 
Among the 136657 transients, 2072 of them met the criterion (selection-A) 
and 1402 of them met the criterion (selection-B), 
resulting in 3474 transients for host galaxy analysis.

\subsection{Identification of Host Galaxies}\label{host}

We looked for host galaxies using Subaru/HSC data
for the 3474 short duration transients selected in the previous section.
The hosts were identified by cross matching transient positions with
the HSC Deep and UltraDeep catalogs in the s16a data release and selecting
galaxies within 1$\arcsec$ radius of the transient. 
Since the HSC astrometry were performed  against with the PS1 PV2 catalog,
we used the PS1 transient positions with the HSC catalogs.
If multiple hosts
were identified, we selected the nearest one. 
We also checked cutout images to exclude noise and bright stars.
The transient's location within the host was computed 
by comparing the difference between the transient's 
coordinates and the host galaxy's coordinates. Multi-detection 
transient coordinates were taken to be the average of the values from individual 
filters, while the host galaxy coordinates were determined in the 
HSC pipeline by comparing and merging peaks from different bands 
using the priority order irzyg (\citealt{hscpipe}).
Out of the 2072 transients meeting the criterion (selection-A) 
and the 1402 transients meeting the criterion (selection-B), 
826 and 301 respectively had identifiable hosts. 
These hosts were then cross-matched with the u*-band and NIR 
catalogs mentioned in $\S$ \ref{uir} to obtain multi-band 
spectral energy distributions (SED) for photometric redshift fitting. 
Not all hosts have complete multi-band data; 
the number of hosts matched with various combinations of bands 
is summarized in Table~\ref{zphot-comb}.
In total, we identified 1127 hosts suitable for fitting.

\begin{table*}
	\caption{Photometric redshift results for PS1 transients.}\label{zphot-comb}
	\begin{center}		
		\begin{tabular}{p{3cm}p{3cm}p{3cm}p{3cm}}\toprule
		\begin{tabular}{r}
		    Combination \tablenotemark{a}\\
		    \\\hline
		    u*(m)+optical+ir \\
		    u*(c)+optical+ir\\
		    u*(m)+optical\\
		    u*(c)+optical\\
		    optical+ir\\
		    optical\\\hline
		    Total
		\end{tabular} &
			\begin{tabular}{rrrr}
				\multicolumn{2}{c}{HSC udeep} \\ \hline
				Total   & zphot $>$ 2 \\ \hline
				295     &   55      \\
				2       &   0       \\
				133     &   52      \\
				12      &   5       \\
				5       &   1       \\
				22      &   9       \\\hline
				469     &   122     
			\end{tabular}
			&
			\begin{tabular}{rrrr}
				\multicolumn{2}{c}{HSC deep}  \\\hline
					Total   &   zphot $>$ 2   \\\hline
					22      &   4           \\
					0       &   0           \\
					97      &   37          \\
					430     &   106         \\
					1       &   0           \\
					108     &   51          \\\hline
					658     &   198         
			\end{tabular}
			&
			\begin{tabular}{rrrr}
        		\multicolumn{2}{c}{All (udeep+deep)}  \\\hline
        			Total   &   zphot $>$ 2   \\\hline
        			317     &   59           \\
        			2       &   0           \\
        			230     &   89          \\
        			442     &   111         \\
        			6       &   1           \\
        			130     &   60          \\\hline
        			1127    &   320 (28$\%$) \tablenotemark{b} \\       
			\end{tabular}\\\hline
			
		\end{tabular}
	\end{center}
	\tablenotetext{a}{u*(m) refers to MUSUBI u*-band, u*(c) refers to CLAUDS
	u*-band, optical refers to HSC s16a data release, and ir refers to UltraVista 
	data release 3.}
    \tablenotetext{b}{The percentage denotes the fraction of transients with
    zphot $>$ 2 relative to the total number of transients}

\end{table*}

\subsection{Photometric Redshift}\label{photometric redshift}

\begin{figure}
\epsscale{1.0}
\plotone{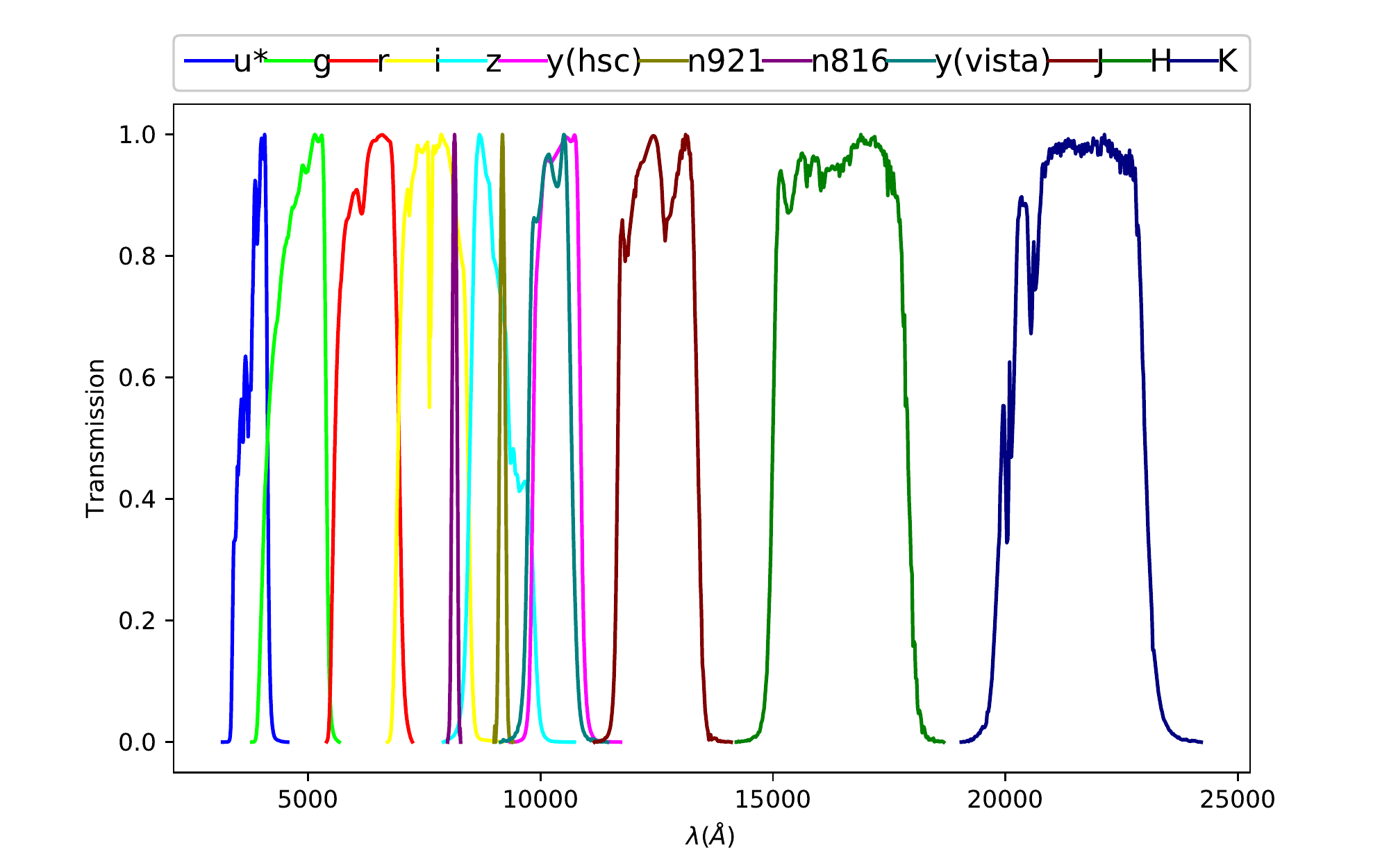} 
\caption{Filter response used for SED fitting. In the legend, y(hsc) refers to HSC y band, while y(vista) refers to UltraVista y band.}\label{filt}
\end{figure}

We used Le Phare to estimate photometric redshifts for the 1127 host galaxies. Le
Phare is a Fortran program that uses the least $\chi^2$ method to
perform SED fitting with templates of stars, quasars and galaxies
(\citealt{lephare1, lephare2}). There are several libraries for galaxy
fitting; instead of choosing a specific library, we merged all the
templates of the default libraries into a single list so that the
fitting could be performed on all the galaxy libraries. In total, there
are 305 templates for galaxy fitting. 
We used filter responses based on each instrument's description. For the U and 
optical bands, we used those given in the filt/cosmos/ directory of the Le Phare 
package: u\_megaprime\_sagem.res, g\_subaru.res, r\_subaru.res, i\_subaru.res, 
z\_subaru.res, WFCAM\_Y.res, NB921.pb, NB816.pb. Since there were no UltraVista 
filter response files in the package, we used the files obtained from the website of 
Peter L. Capak at California Institute of Technology 
\footnote{\url{http://www.astro.caltech.edu/~capak/filters/index.html}}.
The filter response files are plotted in Figure~\ref{filt}.
The cosmological parameters used were: H$_0$ = 70, $\Omega_0$ = 0.3,
$\Lambda_0$ = 0.7. The redshift step used was dz = 0.4, and the maximal
redshift for fitting was set at z$_{max}$ = 6. No extinction laws were
employed in the SED fitting.

We performed a quality assessment of the photometric redsfhits by comparing our 
photometric samples to the spectroscopic redshift catalog in the COSMOS field,
PRIMUS data release 1 (\citealt{primus}).
From the catalog we selected sources with zspec quality =3 or 4 
(where zquality =4 is the highest-quality redshift with
$\sigma_{\delta z}/(1+z) \sim 0.005$ 
and zquality =3 is redshift with $\sigma_{\delta z}/(1+z) \sim 0.022$).

\begin{figure*}
\epsscale{1.0}
\plotone{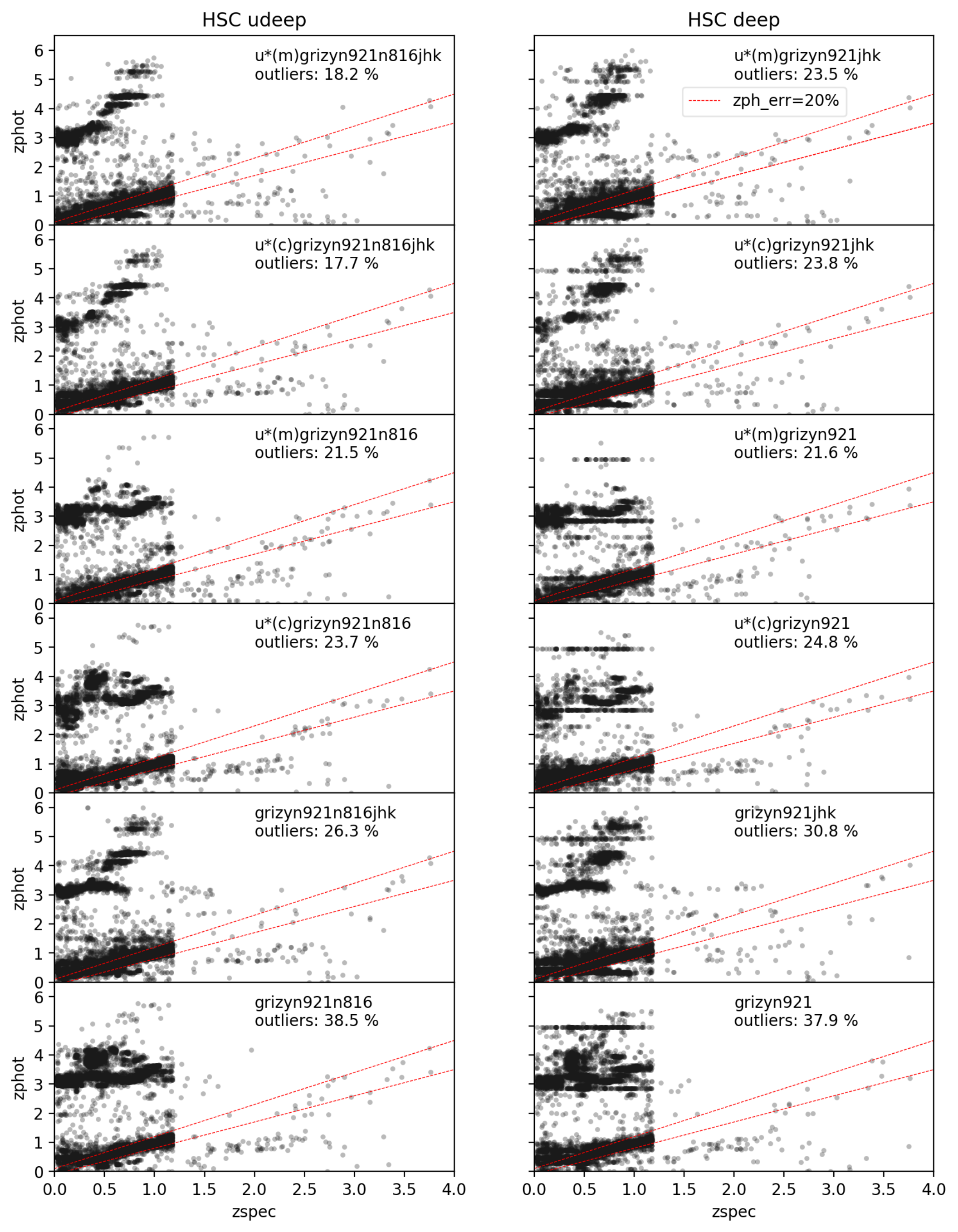} 
\caption{Scatter plot of zphot quality using zspec catalog. Dashed lines are 
zph$\_$err=20\%. Outliers refer to the fraction of zph$\_$err $>$ 20 \%. Zphot: left 
column: HSC udeep (g, r, i, z, y, nb921, nb816 bands), UltraVista (y, j, h, ks 
bands), MUSUBI (u*-band), CLAUDS (u*-band); right column: HSC deep (g, r, i, z, y, 
nb921 bands), UltraVista (y, j, h, ks bands), MUSUBI (u*-band), CLAUDS (u*-band). 
Zspec: G10CosmosCat PRIMUS. u*(c) refers to CLAUDS (u*-band), while u*(m) refers to 
MUSUBI (u*-band)}\label{scat}
\end{figure*}

\begin{figure*}
\epsscale{1.0}
\plotone{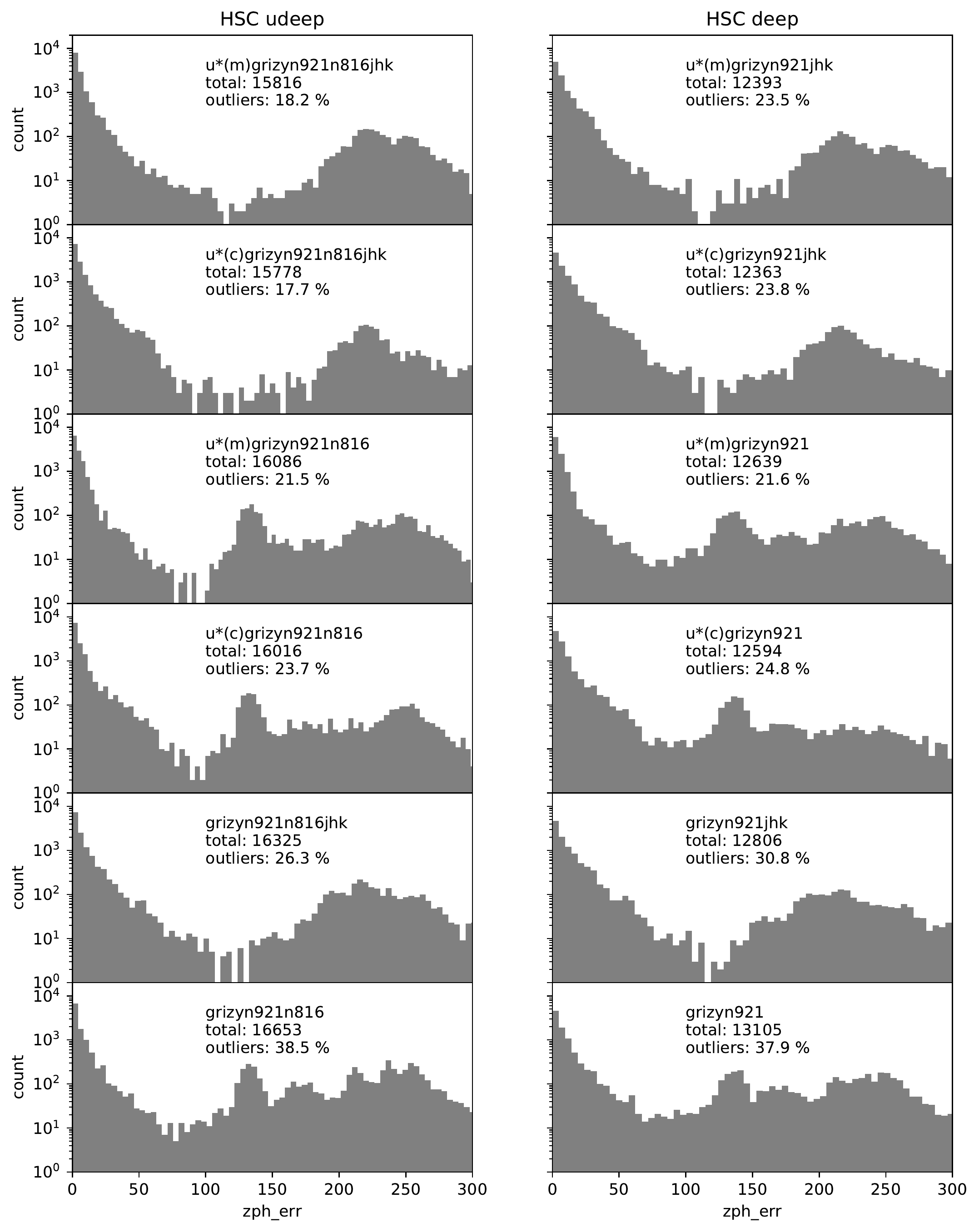}
\caption{Histogram of zphot quality using zspec catalog. Total number of zspec data 
used is specified. Outliers refer to the fraction of zph$\_$err $>$ 20 \%. Zphot: 
left column: HSC udeep (g, r, i, z, y, nb921, nb816 bands), UltraVista (y, j, h, ks 
bands), MUSUBI (u*-band), CLAUDS (u*-band); right column: HSC deep (g, r, i, z, y, 
nb921 bands), UltraVista (y, j, h, ks bands), MUSUBI (u*-band), CLAUDS (u*-band). 
Zspec: G10CosmosCat PRIMUS. u*(c) refers to CLAUDS (u*-band), while u*(m) refers to 
MUSUBI (u*-band)}\label{hist}
\end{figure*}

We assess the photometric redshift quality considering 6 different combinations of 
survey catalogs for each of the two HSC catalogs (Deep and UltraDeep), which are 
listed in Table~\ref{zphot-comb}, because not all host galaxies of our selected 
transients have full SED data coverage.
We chose a 20\% of redshift error ($zph\_err$) as the boundary of outliers,
where $zph\_err=\frac{\left|zspec-zphot\right|}{zspec+1}$. 
Figures~\ref{scat} and~\ref{hist} show the quality of Le Phare on the
PRIMUS catalog and the percentage of outliers. The left and right
columns of the scatter-plot in Figure~\ref{scat}, as well as those of the
histogram in Figure~\ref{hist}, are zphots computed with the HSC s16a udeep
and HSC s16a deep catalogs respectively, as specified above
each column.  
Each row corresponds to each of the 6 combinations of survey catalogs.
The red dashed line in Figure~\ref{scat} is the line where zph$\_$err = 20\%. 
``Total" in Figure~\ref{hist} refers to the number of multi-band data computed. 
``Outliers" refers
to the fraction of those with zph$\_$err $>$ 20\%. We see that by
adding u*-band data, we can reduce the fraction of outliers by 8\%. The
scatter plot shows that most of the data lie within the two dashed
lines. This result indicates that there are significant outliers with zphot$>$2 
that merit additional careful examination for host object classification.

\begin{figure*}
\epsscale{1.0}
\plotone{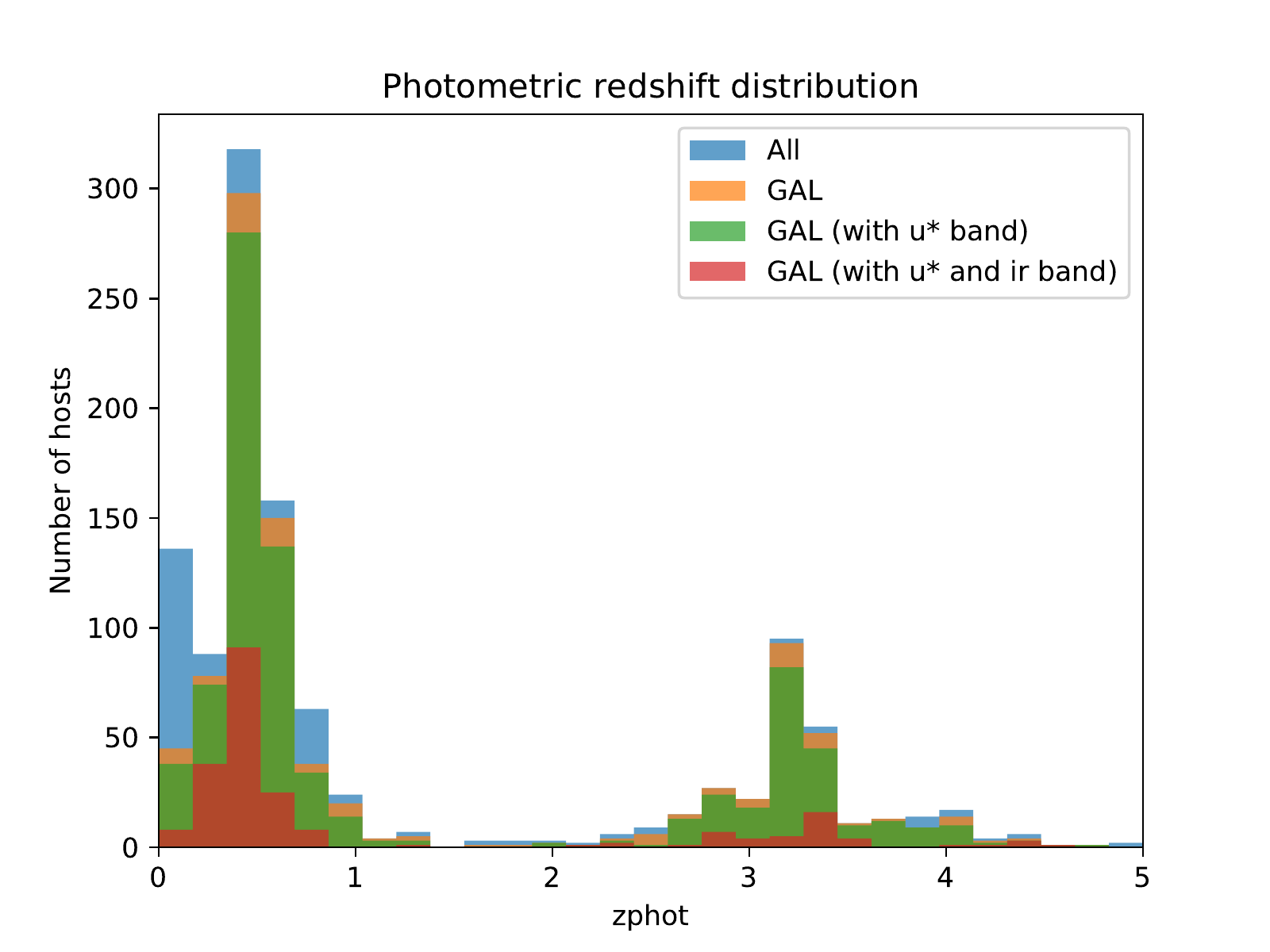} 
\caption{Photometric redshift distribution of 1127 hosts from PS1 transients. Blue 
refers to the Photometric redshift result for all 1127 hosts, which include hosts 
classified as galaxy, QSO, or star in Le Phare. Orange refers to the subset of hosts 
that are classified as galaxies in Le Phare. Green refers to the subset of hosts
that are classified as galaxies and has u*-band data in their SED. Red refers to the 
subset of host that are classified as galaxies and has u*-band and ir data in their 
SED.}\label{zphot-dist}
\end{figure*}

The distribution of the photometric redshifts for our 1127 hosts from PS1 transients 
is shown in Figure~\ref{zphot-dist}. The number of host galaxies with zphot $>$ 2 
from our PS1 transients for various combinations of multi-band SED is shown in 
Table~\ref{zphot-comb}.
The photometric redshifts indicate that about 30\% of our
host galaxy samples fall into the range of zphot $>$2.  
These outlier samples had better SED fits with star or QSO models than with galaxy 
templates. 
We also examined cutout images and found that most of these objects showed
point-like features. 
The relatively small position differences between the transients and these
point-like hosts 
also strongly suggested that these outlier samples might be stars.
Hence, we classified all the outliers as stars or QSOs.

We converted apparent magnitudes of transients to absolute magnitude (M$_{abs}$) 
based on zphot. The M$_{abs}$ is calculated by using the distance module (DM)
calculator\footnote{\url{http://www.astro.ucla.edu/~wright/CosmoCalc.html}}
(\citealt{dmcal}).

\subsection{Known source identification}\label{sec:results}

After an initial examination by eye, we excluded events with bright hosts and flat
or variables (decaying, then rising) light curves. We selected $\sim$ 150 events
with faint hosts or OA-like light curve variations, and performed cross-matching 
against various catalogs using Simbad and VizieR. 
By cross matching, we can exclude the known transients and focus on unidentified 
ones.
We found that most of the transients, whether or not they were zphot outliers,
had star-like or bright hosts. (GRB hosts, by contrast, tend to be faint and small, 
and are of course not star-like).
Out of the $\sim$ 150 events, $\sim$ 20 events had hosts detectable
in X-ray or radio band, which we excluded as probable AGNs.

\begin{figure*}[]
	\begin{center}
		\includegraphics[width=\columnwidth]{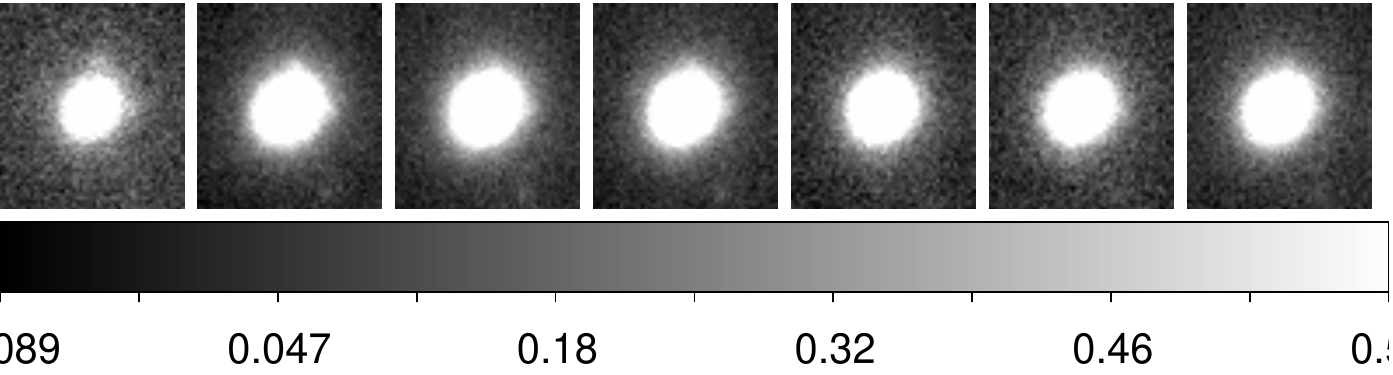}
		\includegraphics[width=\columnwidth]{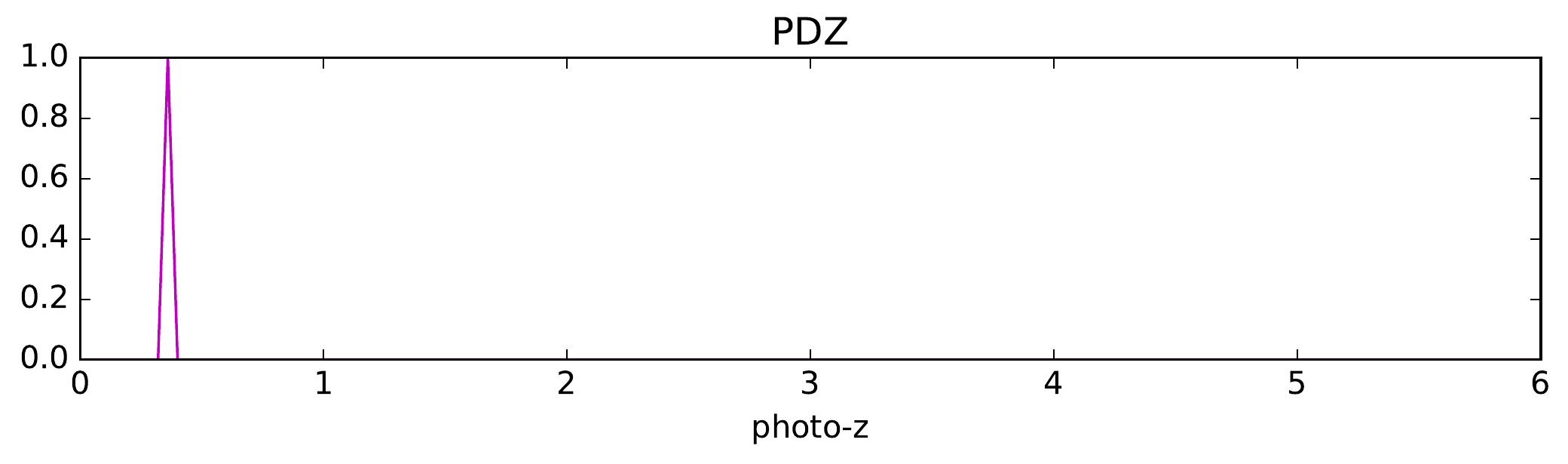}
		\includegraphics[width=\columnwidth]{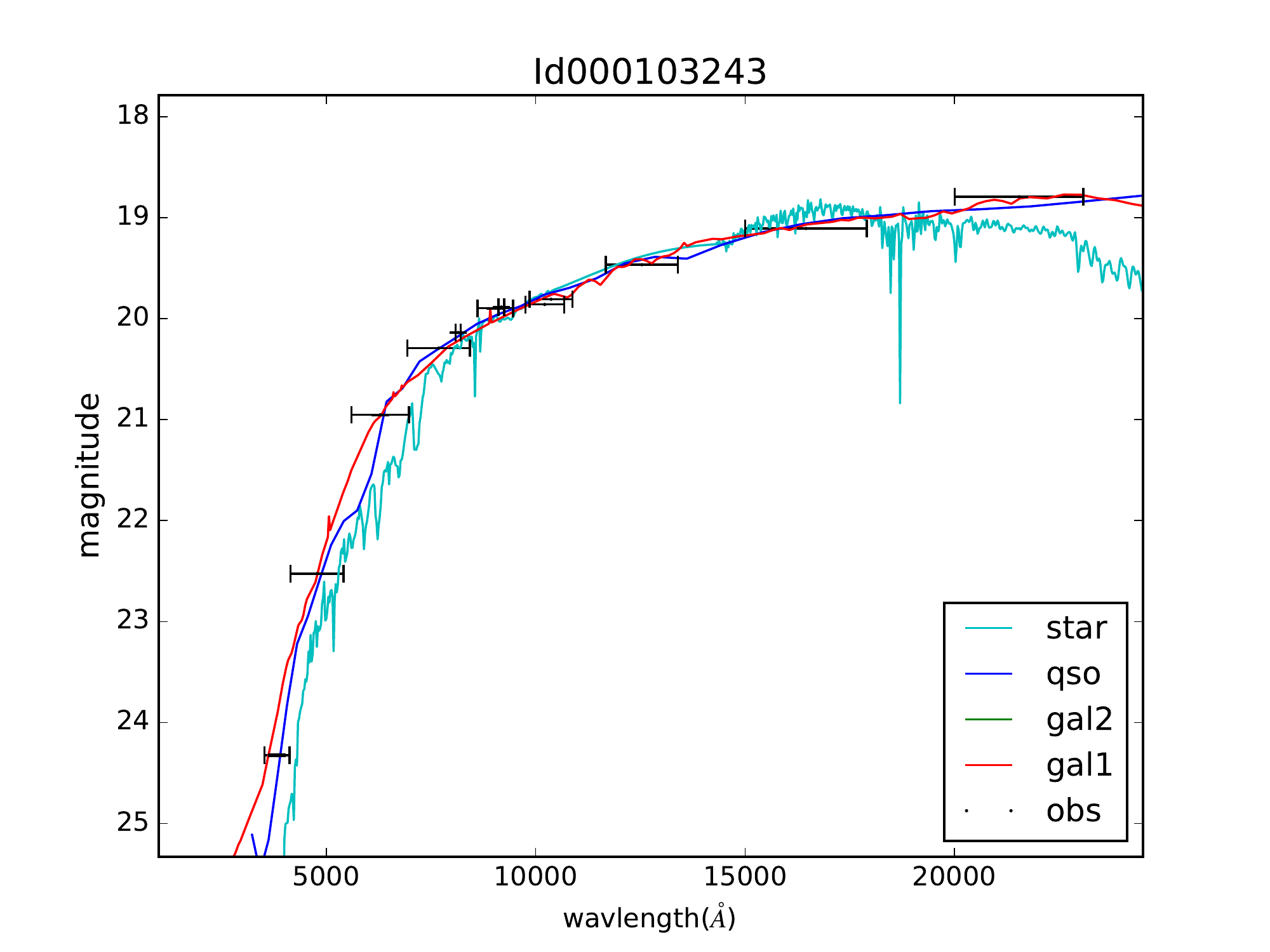}
		\includegraphics[width=\columnwidth]{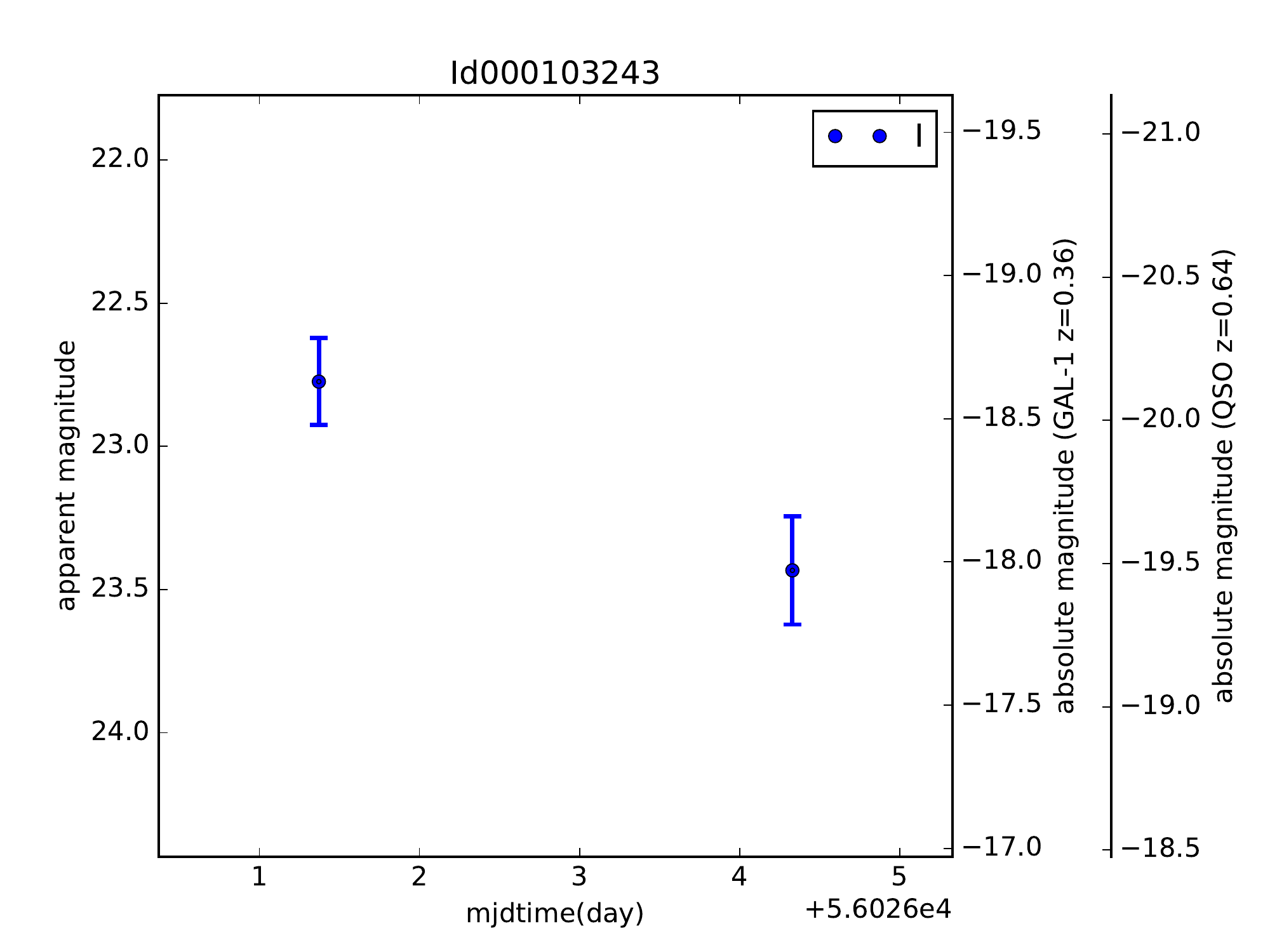}   
		
	\end{center}
	\caption{Example transient. Top-left: Multi-band cutout images of the host 
	galaxy from HSC s16a. From left to right they are g, r, i, z, y, nb816 and nb921 
	band. The transient is located at the center of each cutout image. Bottom-left: 
	SED fitting result of the host galaxy from Le Phare. Gal1 refers to the best 
	fitted galaxy template, while gal2, if there is one, refers to the second best. 
	Top-right: Photo-z probability distribution. Note that this is for galaxy 
	templates only. Bottom-right: Light curve with apparent magnitude on the left 
	scale and absolute magnitude on the right. The absolute magnitudes are 
	calculated using the redshift obtained from Le Phare, which is specified in the 
	parenthesis of the y label on the right. We discuss and conclude that this 
	transient is not an OA but possibly an AGN in $\S$ \ref{sec:selection}.
	}\label{ex}
\end{figure*}

\subsection{Selection of orphan GRB afterglows}\label{sec:selection}

For each of the 1127 transients with hosts identified,
we generated individual event summary including the cutout images of host galaxies, 
zphot result of host galaxies, and light curve of transients with a scale of their 
absolute magnitude;  
we classified the transients by examining this combined information.
We show an example in Figure~\ref{ex} to demonstrate our selection process. This 
transient has an observed duration $\sim$ 3 days and a decaying light
curve. Multi-band data from u*-band, optical, and IR give us a zphot of 0.36.
The cutout images show that the host is very bright, 
and thus unlikely to be an OA host. 
From cross matching catalogs, we find in Simbad that it is a galaxy in the Advanced 
Camera for Surveys - General Catalog (ACS- GC)
with zspec=0.3396, which is highly consistent with our zphot result.
It also has detections in the Chandra and XMM-Newton surveys.
As a result, we conclude that it is unlikely to be an OA, but could be an AGN.

After careful examinations, we concluded that no OA candidates were found. 
Most of the transients had too few light curve points to show any 
significant properties, and non of them met the duration selection criteria.
The light curves of those with more data points were usually either very flat, 
or had variation amplitudes that were too small ($<$ 1 magnitude) 
when compared to the theoretical OA light curves. 
From the cut out images and SED, most host galaxies appeared to be very bright 
and large, while the expected host properties of OAs are dwarf galaxies, 
with relatively fainter magnitudes and redshifts around 1 $\sim$ 2. 
We were not able to find a candidate with both an interesting light curve 
and host properties strongly indicative of being an OA. 
We summarize our selection process and result in Table~\ref{flow}.

\begin{table*}
	\caption{Summary of number of candidates remaining after each selection 
	cut.}\label{flow}
	\begin{center}	
		\begin{tabular}{ p{7cm}|@{}>{\centering\arraybackslash}p{5.3cm}@{} }\toprule
			Selection Criteria & Number\\\hline
			Database of differential images from PS1 MD04 & 136657\\\hline
			Duration $t$, count \& light curve  &
			\begin{tabular}{>{\centering\arraybackslash}p{2.3cm}|
			>{\centering\arraybackslash}p{2.3cm}}
			$\geq$ 3 detections within 15 days &
			= 2 detections within 4 days \& decaying \\\hline
			2072 & 1402\\
		    \end{tabular}\\\hline
			Host in HSC          & 
			\begin{tabular}[b]{>{\centering\arraybackslash}p{2.3cm}|
			>{\centering\arraybackslash}p{2.3cm}}
			826 & 301\\
			\end{tabular}\\\hline
		    Examine host galaxies, light curve properties,
		    zphots \& cross matching catalogs & 0\\\hline
		\end{tabular}
	\end{center}
\end{table*}

\section{Discussion}\label{sec:discussion}
\subsection{Comparison with theoretical expectations}\label{sec:theory}
The theoretically predicted rate of OAs depends on the model used. 
Here we compute and compare with our null result the expected rate in the
PS1 MD04 field using models from three different theoretical
papers:~\cite{T02} (T02), 
~\cite{R08} (R08), and ~\cite{N02} (N02). 
We assume a survey area $\Omega_{obs}$ of 7 deg$^{2}$,
and a limiting magnitude in R band $\sim$ 23. 
We use 154 nights with 1-13 nights' separation and a total observing time 
of 320 nights to compute the number of expected OAs, 
following the respective calculation methods in each of the three papers.
In the following calculations, N$_{oa}$ is the number of OAs expected to be
observed in our survey; N$_{snap}$ is the number of OAs in one snapshot of the
whole sky; T$_{obs}$ is the total observing time (320 nights); 
T$_{oa}$ is the lifetime of an OA. To identify a transient,
we need to consider the consecutive monitoring condition,
which we define as the density of the log file shown in Figure~\ref{log}.
In our rate calculations we define a survey efficiency $eff$ equal to this
density, which is 154/320 $\sim$ 0.5. 
 
T02 used GRB parameters from an average of 10 well studied events.
From this model and Table 1 of T02, we found that N$_{snap} = $ 330 can 
reproduce the expected 36 OAs with a sensitivity of R $\sim$ 23 and 
effective survey area of about 4500 deg$^2$. We use N$_{snap} = $ 330 and the
$\langle1/T_{oa}\rangle^{-1} \sim$ 18 days from Table 1 of T02 for SDSS to
estimate detection rate with our cadence. 
\begin{eqnarray}
N_{oa}&=&N_{snap}\cdot\frac{equivalent\ snapshot\ area}{4\pi}\\
&=&N_{snap}\cdot\frac{T_{obs}}{\langle1/T_{oa}\rangle^{-1}}
\cdot\frac{\Omega_{obs}}{4\pi}\cdot eff\\
&=&330\cdot\frac{320\ day}{18\ day}\cdot\frac{7\ deg^{2}}{4\pi}
\cdot\left(\frac{\pi}{180\ deg}\right)^{2}\cdot 0.5\\
&\simeq&0.5,
\end{eqnarray}
which is smaller than one but not exactly zero.
However, the OAs considered by T02 are bright, yielding average lifetimes that are
too long ($\langle1/T\rangle^{-1} \sim$ 18). If we instead considered an average
lifetime of about a week at this sensitivity \citep[e.g.][]{kann},
we would obtain an expected number 
of 1 to 2 OAs, which is larger than our null result. 

R08 considered the universal structured jet (USJ) model, which has a wide 
outflow $\theta_{jet}$ = 90$^{\circ}$ and an angle dependent energy distribution
E($\theta$) $\propto \theta^{-2}$. 
They suggested that if the consecutive monitoring T$_{obs}$ is longer than the OA 
mean lifetime T$_{th}$, then the number of OAs is
\begin{equation}
N_{oa} = R_{oa}\cdot T_{obs}\cdot\frac{\Omega_{obs}}{4\pi},
\end{equation}
where R$_{oa}$ is the mean rate that OA appear in the sky over the survey 
flux threshold.
At R $\sim$ 23, according to the calculations for the expected rate of the
survey conducted by \cite{rau2006} in R08, T$_{th} \sim$ 22 and 
R$_{oa} \sim$ 6.3 day$^{-1}$. Putting these parameters as well as the survey
efficiency $eff$ into equation (6) we obtain
\begin{equation}
N_{oa} = 6.3\ day^{-1}\cdot 320\ day\cdot\frac{7\ deg^2}{4\pi}
\cdot\left(\frac{\pi}{180\ deg}\right)^{2}\cdot0.5\simeq0.2.
\end{equation}

N02 considered a jet with constant maximal OA observing angle $\theta_{max}$ 
that is independent of $\theta_{jet}$ (for $\theta_{jet} < \theta_{max}$), 
with two jet opening angles.
The ``canonical" model (N02C) used $\theta_{jet}$ = 0.1 radian,
which is similar to the averaged jet opening angle.
The ``optimistic" model (N02O) used $\theta_{jet}$ = 0.05 radian, which is even 
smaller than observed typical jet opening angle \citep[e.g.][]{racusin09}.
They did not give a characteristic lifetime of OAs, but stated that for most OAs,
if the separation between observed nights is longer than 2 weeks, then the two 
observations can be considered independent. 
We obtain N$_{snap}$ values from Figure 3 of N02, 
which is also supported by Figure 8 of R08. 
Using T$_{oa}$ $\sim$ 14, N$_{snap} \sim$ 70 (N02O) and 
N$_{snap} \sim$ 1.7 (N02C), we find the estimated number of OAs to be:
\begin{eqnarray}
	N_{oa}&=&N_{snap}\cdot\frac{T_{obs}}{T_{oa}}\cdot\frac{\Omega_{obs}}{4\pi}
	\cdot eff\\
	&=&70\cdot\frac{320\ day}{14\ day}\cdot\frac{7\ deg^{2}}{4\pi}
	\cdot\left(\frac{\pi}{180\ deg}\right)^{2}\cdot 0.5\\
	&\simeq&0.1\  (N02O)\\
	N_{oa}&=&1.7\cdot\frac{320\ day}{14\ day}\cdot\frac{7\ deg^{2}}{4\pi}
	\cdot\left(\frac{\pi}{180\ deg}\right)^{2}\cdot 0.5\\
	&\simeq&0.003\  (N02C).
\end{eqnarray}
Table~\ref{rate} shows N$_{oa}$ calculated by the above models with
the given parameters.

\begin{table*}
	\caption{Predicted number of OAs in PS1 MD04 using different models.}\label{rate}

	\begin{center}
		\begin{tabular}{rrrrrD}\toprule
			Model & 10$^{\langle\log_{10} T\rangle}$ & $\langle1/T\rangle^{-1}$ & N$_{snap}$ & N$_{oa}$ & N$_{oa}$ \\
			& (day)\tablenotemark{a} & (day) & (all sky) & & (one week lifetime)\tablenotemark{b}\\\hline
			T02 & 32 & 18 & 330 & 0.5  & 1.3 \\
			R08 & 22 & 2  & 12  & 0.2  & 0.05\\
			N02O& ...& ... & 70  & 0.1 & 0.3 \\
			N02C& ...& ... & 1.7 & 0.003 & 0.006\\\hline
		\end{tabular}
	\end{center}
	\tablecomments{N02 did not give the lifetimes 10$^{\langle\log_{10} T\rangle}$ 
	and $\langle1/T\rangle^{-1}$, so they are left blank. But they mentioned that
	two weeks can be considered as an average lifetime of OAs. Therefore, we used
	T $\sim$ 14 for their N$_{oa}$ calculation.}
	\tablenotetext{a}{$T_{th}$ in R08.}
	\tablenotetext{b}{Calculated with a shorter lifetime (one week) than
	the lifetime originally considered in the models.}
\end{table*}

In summary, R08 predicts 0.2 OAs, which we find consistent with our result. 
Even though R08 considered a very
short lifetime $\langle1/T\rangle^{-1} \sim$ 2, the predicted number of OAs would
be still closer to zero (N$_{oa} \sim$ 0.05) if we increase the lifetime to a
week. Thus, we find R08 consistent with our result in this case also.
N02O predicts 0.1 OAs considering an average lifetime of two weeks.
Reducing the lifetime to one week would increase the N$_{oa}$ to 0.2,
which we find consistent with our result. N02C, the most pessimistic of the models
we considered, predicts 0.003 OAs; this, too, is consistent with our null result. 

\subsection{Toward further surveys}
We next discuss methods to evaluate the rate predictions of the models
R08 and N02. If we were to examine all 10 fields of the PS1 Medium Deep Survey,
R08 and N02O predict that we would find 1 or 2 OAs, each of which could 
be further evaluated. But N02C predicts 0.03 OAs, indicating that we would be
still unlikely to find an OA in this survey. 
With the same survey area (7deg$^{2}\times$10), number of observing nights,
and survey efficiency, N02C requires a limiting magnitude of R$\sim$25
to detect 1 OA. Therefore, we will need a more efficient survey to evaluate N02C.

From the above calculations, we can see that N$_{snap}$ varies greatly 
from model to model. However, $\langle1/T_{oa}\rangle^{-1}$ also varies
by roughly the same order, leading to same-order values of N$_{oa}$s.
To distinguish between the various models, we find continuous monitoring for a 
week each month for a year is preferred over observing 84 nights continuously,
since at a sensitivity of R $\sim$ 23 we can regard the monthly survey as 12
snapshots and would not have to consider the various characteristic lifetimes. 
Depending on the model, the monthly survey might produce a higher N$_{oa}$ than
the continuous 84 nights survey, but it might also increase the difficulty of 
classifying transients.

As of its second data release (dr2),
HSC has observed 174 nights (\citealt{hscdr2}).
The Deep survey monitored four fields,
amounting to a total area of $\sim$ 26 deg$^2$. 
The cadence for each field each year obtained from dr2
is roughly two weeks per month 
over the course of two to four consecutive months. 
The Ultra Deep survey observed two fields, with a total area of $\sim$ 4 deg$^2$. 
The cadence for each field is roughly one to two weeks every one or two months over 
the course of half a year. Using the aforementioned cadence of HSC dr2 and the four 
models described in $\S$ \ref{sec:theory}, the predicted number of OAs and the 
parameters used for the calculations are summarized in Table~\ref{hsc-lsst}. Note 
that $\langle T_{obs}\rangle$ for HSC in Table~\ref{hsc-lsst} shows the yearly 
effective survey duration averaged over the multiple fields, which is only a rough 
estimate of the actual effective survey duration used for the calculations.
The HSC survey would be able to verify the N02O case as summarized in 
Table~\ref{hsc-lsst}.
LSST will image 10,000 deg$^{2}$ every three nights, 
with a limiting magnitude of R $\sim$ 24.5
(\citealt{ivezic2019, lsst2017}).
We used these survey parameters of LSST to obtain the expected numbers of OAs in 
Table~\ref{hsc-lsst}.
LSST would be able to verify all four model cases in Table~\ref{hsc-lsst}.
Our expected numbers of OAs for T02 and N02C are in the same order as estimations by
\citet{2009arXiv0912.0201L} (i.e. N$_{oa}\sim1000$, which used the T02 model) and  by
\citet{ghirlanda2015} (i.e. N$_{oa}\sim$ 50, which used a population
synthesis code), respectively.
We also calculated the expected numbers of OAs with the planned supernova survey 
using the WFIRST mission \citep{2013arXiv1305.5425S}. 
In Table 4, we summarized the results for three layers of SN surveys (wide, medium, 
and deep) with the limiting magnitude of J-band, the cadence of 5 days, and the 
planned survey duration of 0.5 years in a 2-year interval.
Since the satellite based time-domain survey would maintain the planned cadence
(i.e. unaffected by weather condition unlike ground based observations), the WFIRST 
survey would also be essential for the OA surveys. Subaru/HSC, LSST and WFIRST 
therefore are promising surveys for discovering OAs.

\setlength\tabcolsep{3pt}
\begin{table*}
\scriptsize
	\caption{Predicted number of OAs for HSC, LSST, and WFIRST using different 
	models. 
	}\label{hsc-lsst}
	\begin{center}		
	
		\begin{tabular}{p{4.6cm}p{2.2cm}p{1.5cm}p{1.4cm}p{3.6cm}p{3.2cm}}\toprule
			\begin{tabular}{p{2.2cm}p{0.9cm}r}
				Survey    & Mag &        Area \\
				          &     & (deg$^{2}$)  \\
				          &         &           \\\hline
				HSC Deep  &  R: 26 &          26  \\
				HSC UltraDeep &  R:  27 &          4   \\
				LSST      &  R: 24.5 &       3300  \\
				WFIRST Wide & J: 27.5 & 27.44 \\
				WFIRST Medium & J: 27.6 & 8.96 \\
				WFIRST Deep & J: 29.3 & 5.04
			\end{tabular}
			&
			\begin{tabular}{rrr}
				\multicolumn{3}{c}{$\langle T_{obs}\rangle$} \\
				\multicolumn{3}{c}{(day yr$^{-1}$)\tablenotemark{a}}\\\hline
				  T02 & R08 & N02  \\ \hline
				  96 & 103&  22 \\
				  101 & 263&  25 \\
				   365 & 365& 365 \\
				   91 & 91& 91\\
				   91 & 91& 91\\
				   91 & 91& 91
			\end{tabular}
			&
			\begin{tabular}{c}
				\multicolumn{1}{c}{10$^{\langle\log_{10} T\rangle}$} \\ 
				\multicolumn{1}{c}{(day)\tablenotemark{b}}\\\hline
                    R08  \\ \hline
				  170\\
				  300\\
				   60\\
				   500\\
				   500\\
				   1200
			\end{tabular}
			&
			\begin{tabular}{rr}
				\multicolumn{2}{c}{$\langle1/T\rangle^{-1}$} \\ 
				\multicolumn{2}{c}{(day)\tablenotemark{c}}\\\hline
				  T02 & R08  \\ \hline
				  97 &  22\\
				  180 &  70\\
				  40 &   8  \\
				  300& 150\\
				  300& 150\\
				  700& 600
				  
			\end{tabular}
			&
			\begin{tabular}{rrrr}
				\multicolumn{4}{c}{N$_{snap}$} \\ 
				\multicolumn{4}{c}{(all sky)} \\\hline
				  T02 & R08 & N02O & N02C \\ \hline
				 5000 & 200 & 1000 &  40 \\
				15000 & 450 & 2000 &  80 \\
				 1200 &  55 &  250 &   8 \\
				 50000 & 1200& 6000& 300\\
				 50000 & 1200& 6000& 300\\
				 170000 & 4000& 30000& 2000\\
			\end{tabular}
			&
			\begin{tabular}{rrrr}
					\multicolumn{4}{c}{N$_{oa}$}  \\
					\multicolumn{4}{c}{(yr$^{-1}$)}  \\\hline
						T02& R08 & N02O & N02C \\\hline
						4 & 0.2 & 2    & 0.07  \\
						1  & 0.1 & 0.5  & 0.02 \\
						885 & 203 & 527  & 17   \\
						17 & 0.4& 26 & 1.3\\
						5 &  0.1& 9& 0.4\\
						10 & 0.2& 24 & 1.6
			\end{tabular}\\\hline
		\end{tabular}
	\end{center}
		\tablecomments{For estimations of HSC surveys, we use the cadence obtained 
		from the HSC second data release (\citealt{hscdr2}). LSST will cover 10000 
		deg$^{2}$ every three days. The three layers of WFIRST SN Survey (wide, 
		medium, deep) will have a survey duration of 0.5 years over a 2-year 
		interval, and a cadence of 5 days \citep{2013arXiv1305.5425S}.}
		\tablenotetext{a}{For HSC surveys this is the yearly effective survey 
		duration averaged over the multiple fields, which is a rough estimate of the 
		actual effective survey duration used for the calculations.}
		\tablenotetext{b}{$T_{th}$ in R08.}
		\tablenotetext{c}{For N02 models we used T $\sim$ 14, same as that in 
		Table~\ref{rate}.}

\end{table*}

\section{Conclusion}\label{sec:conclusion}
In an attempt to find OAs, we used the MD04 field of the Pan-STARRS1 
Medium Deep Survey, which covers an area of 7 deg$^{2}$ overlapping 
with the COSMOS field, has a limiting magnitude reaching R $\sim$ 23,
and observed 154 nights over the course of 2 years
(from December 2011 to January 2014).
We identified 136657 transients by generating
differential images, and then performed transient classification.
We reduced the number to 1127 by excluding long duration and hostless transients,
and then checked carefully each remaining candidate's host galaxy,
location in host, light curve properties and zphot, cross-matching the results
with other catalogs. For zphot, we used MUSUBI, CLAUDS, HSC and UltraVista to
construct the SED of each transient, and then performed SED fitting using
Le Phare. We checked our zphot quality using a zspec catalog PRIMUS and found
that we could reduce the fraction of outliers (zph\_err $>$ 20\%) down to 18\%,
which we consider acceptable. After careful examination we concluded that
we did not find any OA candidates. 

We then compared our result with the expected number of OAs computed using
different models: T02, R08, N02O and N02C which predicts 0.5, 0.2, 0.1 and
0.003 OAs, respectively. R08 and N02 are consistent with our result. 
We find the average lifetime of OAs in T02 is too long
($\langle1/T_{oa}\rangle^{-1} \sim$ 18); reducing it would increase N$_{oa}$
to 1 $\sim$ 2, which we consider to be marginally consistent with our result.
We may be able to evaluate R08 and N02O by examining all 10 fields of
PS1 Medium Deep Survey, but the evaluation of N02C would require a more
efficient survey.

HSC is 3 $\sim$ 4 magnitudes deeper than PS1, which should result in a
higher rate of OA candidate detections, and probably in a discovery. 
However, LSST is the most promising survey to detect OAs, and will 
detect a larger number than other optical surveys.

\acknowledgments
%
The Pan-STARRS1 Surveys have been made possible through contributions of the 
Institute for Astronomy, the University of Hawaii, the Pan-STARRS Project Office,
the Max-Planck Society and its participating institutes, the Max Planck Institute
for Astronomy, Heidelberg and the Max Planck Institute for Extraterrestrial Physics, 
Garching, The Johns Hopkins University, Durham University, the University of 
Edinburgh, Queen’s University Belfast, the Harvard- Smithsonian Center for 
Astrophysics, the Las Cumbres Observatory Global Telescope Network Incorporated, the 
National Central University of Taiwan, the Space Telescope Science Institute, the 
National Aeronautics and Space Administration under Grant No. NNX08AR22G issued 
through the Planetary Science Division of the NASA Science Mission Directorate, the 
National Science Foundation under Grant No. AST-1238877, the University of Maryland, 
Eotvos Lorand University (ELTE), and the Los Alamos National Laboratory.

The Hyper Suprime-Cam (HSC) collaboration includes the astronomical communities of 
Japan and Taiwan, and Princeton University.  The HSC instrumentation and software 
were developed by the National Astronomical Observatory of Japan (NAOJ), the Kavli 
Institute for the Physics and Mathematics of the Universe (Kavli IPMU), the 
University of Tokyo, the High Energy Accelerator Research Organization (KEK), the 
Academia Sinica Institute for Astronomy and Astrophysics in Taiwan (ASIAA), and 
Princeton University.  Funding was contributed by the FIRST program from Japanese 
Cabinet Office, the Ministry of Education, Culture, Sports, Science and Technology 
(MEXT), the Japan Society for the Promotion of Science (JSPS),  Japan Science and 
Technology Agency  (JST),  the Toray Science  Foundation, NAOJ, Kavli IPMU, KEK, 
ASIAA,  and Princeton University.
Based the HSC-SSP on data collected at the Subaru Telescope and retrieved from the 
HSC data archive system, which is operated by the Subaru Telescope and Astronomy
Data Center at National Astronomical Observatory of Japan.

This paper makes use of software developed for the Large Synoptic Survey Telescope. 
We thank the LSST Project for making their code available as free software at 
http://dm.lsst.org.

This work is supported by the Ministry of Science and Technology of Taiwan grants 
MOST 105-2112-M-008-013-MY3 (Y.U.) and MOST 108-2112-M-001-051 (K.A.).

\end{document}